\documentclass[prl,aps,10pt, floatfix, twocolumn]{revtex4}
\usepackage{inputenc,amsmath,amssymb}
\usepackage{times}
\usepackage{graphicx}
\usepackage{epsfig}
\usepackage{color}
\begin{document}

\title{Uncovering a Nonclassicality of the Schr\"{o}dinger Coherent State up to the Macro-Domain}
\author{S. Bose$^{1}$, D. Home$^2$ and S. Mal$^3$}
\affiliation{$^1$Department of Physics and Astronomy, University
College London, Gower St., London WC1E 6BT, UK}
\affiliation{$^2$CAPSS, Physics Department, Bose Institute, Salt Lake, Sector V, Kolkata
700097, India}
\affiliation{$^3$ S.N.Bose National Center for Basic Sciences, Block JD, Sector III, Salt Lake, Kolkata-700098, India}
\date{\today}


\begin{abstract}
The Leggett-Garg inequality (LGI), based on the notions of realism and noninvasive measurability, is applied in the context of a linear harmonic oscillator.  It is found that merely through observing  at various instants which region of the potential well, the oscillating quantum object is in, the LGI can be violated without taking recourse to any ancillary quantum system. Strikingly, this violation reveals an unexplored nonclassicality of the state which is considered the most ``classical-like" of all quantum states, namely the Schr\"{o}dinger coherent state. In the macrolimit, the extent to which such nonclassicality persists for large values of mass and classical amplitudes of oscillation is quantitatively investigated. It is found that while for any given mass and oscillator frequency, a significant quantum violation of LGI can be obtained by suitably choosing the initial peak momentum of the coherent state wave packet, as the mass is sufficiently increased, actual observability of this violation becomes increasingly difficult. A feasible experimental setup for testing the predicted quantum mechanical violation of LGI is suggested using a trapped nano-object of $\sim 10^6-10^9$ amu mass.

\end{abstract}

\maketitle

\textit{Introduction}: Central to the classical world-view is the basic notion of realism, viz. that at any instant, irrespective of any measurement, a system is in a \textit{definite} one of the available states for which all its observable properties have definite values. A stimulating direction for probing  the quantum mechanical (QM) incompatibility with the notion of realism is provided by the Leggett-Garg inequality (LGI) \cite{leggett,leggett1,leggett2}. LGI is formulated in terms of time-separated correlation functions corresponding to successive measurement outcomes for a system. Apart from the idea of \textit{realism}, a necessary ingredient for obtaining LGI is the notion of \textit{noninvasive measurability} (NIM) which implies that it is possible to determine which of the states the system is in, without affecting the system's subsequent behaviour. Experimental verification of the QM violation of LGI would, therefore, signify repudiation of the notion of realism that includes the assumption of NIM. Here it is important to note that NIM can be regarded a ``natural corollary'' of the assumption of realism, as Leggett \cite{leggett, leggett1, leggett2} has argued, by considering what may be called the `\textit{negative result measurement}' (NRM) procedure (this will be explained shortly). Thus, if the NRM procedure is implemented loophole-free, the testing of LGI would enable a clearer scrutiny of the notion of realism.

While the original motivation that led to LGI was for testing the possible limits of QM in the macroscopic regime, e.g., in terms of suitable experiments involving the rf-SQUID device \cite{van}, in recent years, there has been a variety of studies (reviewed, for example, by Emary et al. \cite{emary}), bringing out various fundamental implications of LGI \cite{kofler}, as well as exploring its aspects pertaining to systems, ranging from, say, solid-state qubits \cite{ruskov,knee}, nuclear spins \cite{athalye}, photons \cite{gossin}, electrons \cite{emary1} to oscillating kaons and neutrinos \cite{home}. Recent suggestions include qubit-oscillator hybrid systems \cite{Rabl} and quantum walks of an atom in a lattice \cite{rob}.

Against this backdrop, the present paper points out and studies an application of LGI to the archetypal example of a linear harmonic oscillator (LHO) which has well defined classical and quantum descriptions. The systems used so far for probing LGI have essentially been qubits, or systems isomorphic to qubits. The only exception, very recent, is the paper by Asadian et. al. \cite{Rabl}, in which a harmonic oscillator is coupled to a qubit and probed through it. In contrast, the LHO example we consider, while involving continuous variables, does not need to couple the oscillator to any auxillary quantum system or degree of freedom. Therefore, to apply LGI here, discretization is needed which is ensured by considering coarse-grained measurement of a type that would determine \textit{which} one of the halves of the region, the oscillating particle is \textit{in} at any given instant, without providing any further information about the position of the particle. This type of measurement is similar to the kind of spatial measurement used in a recent realization of the violation of LGI in a multiple coupled well structure \cite{rob}. Invoking such dichotomic measurements, it turns out that the LHO example serves to demonstrate the power of LGI in revealing a testable non-classical feature of the Schr\"{o}dinger coherent state (non-spreading wave packet with minimum position-momentum uncertainty product)\cite{schrodinger} whose quantum dynamical behaviour is similar to that of a classical oscillator and is regarded as providing the best possible classical-like quantum description of LHO. Using this coherent state, the extent to which for even larger values of mass,  the QM violation of LGI persists is investigated. A number of key features of the calculated results are highlighted, and the actual feasibility of a relevant experimental test involving nano-objects is discussed. We proceed by first setting up the relevant form of LGI and explain in the context of our example how the condition of NIM underlying LGI can be satisfied by using NRM. Here it is worth noting that although a number of experiments have tested LGI, only two to date \cite{knee, rob} have claimed to have satisfied the condition of NIM through ideal NRM. Hence it is desirable to have further tests of LGI by unambiguously satisfying the condition of NRM. 

\textit{LGI and the notion of NRM}: In the one-dimensional LHO example considered in this paper, the temporal evolution involves oscillation between two states, one of which corresponds to the particle being found within, say, the negative half of the region ($x=0$ to $x\rightarrow -\infty$) which we call the state 1, while the state 2 pertains to the particle being found within the positive half ($x=0$ to $x\rightarrow +\infty$). Let Q(t) be an observable quantity such that at any instant, it takes a value $+1(-1)$ depending on whether the system is in the state 1(2). Now, consider a set of runs starting from the identical initial state such that on the first subset Q is measured at times $t_{1}$ and $t_{2}$, on the second at $t_{2}$ and $t_{3}$, on the third at $t_{3}$ and $t_{4}$, and on the fourth at $t_{1}$ and $t_{4}$ (here $t_{1}<t_{2}<t_{3}<t_{4}$). From such measurements, one can obtain the temporal correlations $C_{ij} \equiv \langle Q(t_{i})Q(t_{j})\rangle$. Then, adapting in this context, the standard argument leading to a Bell-type inequality with the measurement times $t_{i}$ playing the role of apparatus settings, the following consequence of the assumptions of realism and NIM is invoked. For sets of runs corresponding to the same initial state, an individual $Q(t_i)$ is taken to have the \textit{same} definite value(+1 or -1), irrespective of the pair \textbf{$Q(t_{i}) Q(t_{j})$} in which it occurs; i.e., the value of $Q(t_i)$ in any pair does \textit{not} depend on whether any prior measurement has been made on the system. Consequently, the combination $[Q(t_{1})Q(t_{2}) + Q(t_{2})Q(t_{3}) + Q(t_{3})Q(t_{4}) - Q(t_{1})Q(t_{4})]$ is always +2 or $-2$. If all these product terms are replaced by their respective averages over the entire ensemble of runs, assuming the principle of induction, the following form of LGI is then obtained

\begin{equation}
\label{LG1}
C \equiv C_{12} + C_{23} + C_{34} - C_{14} \leq 2.
\end{equation}

\noindent The above is, thus, a testable inequality imposing realist constraints on the time-separated correlation functions. Now, to explain how the notion of NIM can be satisfied by invoking NRM, let us consider the case in which Q is measured at $t_{1}$, followed by at $t_{2}$, corresponding to the determination of the correlation function $C_{12} = P_{++}(t_{1}, t_{2})- P_{+-}(t_{1}, t_{2})+P_{--}(t_{1}, t_{2}) - P_{-+}(t_{1}, t_{2}) $ where $P_{++}(t_{1}, t_{2})$ is the joint probability of finding the particle in the state 1 at both the instants $t_{1}$ and $t_{2}$; similarly, for $P_{+-}(t_{1}, t_{2}), P_{--}(t_{1}, t_{2}), P_{-+}(t_{1}, t_{2})$. Note that the derivation of LGI requires essentially the first measurement of each such pair to satisfy NIM. This can be ensured through the NRM procedure by arranging the measuring setup so that if, say, the probe is triggered, one can infer $Q(t_{1}) = +1$, while if it is \textit{not}, $Q(t_{1}) = -1$, thereby the probe being untriggered provides information about the value of Q=-1, although there is \textit{no interaction} occurring between the probe and the measured particle; NIM is, thus, satisfied. Now, if the results of those runs are only used for which $Q(t_{1}) = -1$, followed by the measurement of Q at $t_{2}$, discarding the results of the rest runs, these results can be used for determining the joint probabilities $P_{-+}(t_{1}, t_{2})$ and $P_{--}(t_{1}, t_{2})$. Similarly, for determining the other two joint probabilities $P_{+-}(t_{1}, t_{2})$ and $P_{++}(t_{1}, t_{2})$, the  measuring setup can be inverted. In this way, one can determine all the 2-time correlation functions occurring in the LGI by ensuring NIM (using NRM) for the first measurement of any pair. The violation of LGI thus obtained would then repudiate the notion of realism because, as Leggett \cite{leggett, leggett1, leggett2} has argued, the `realist' statement that the particle `has' a definite state at any instant is hard to justify if the state can be affected by the NRM procedure. It is, therefore, necessary to invoke the NRM procedure in order to ensure NIM for achieving loophole-free verification of LGI that can be regarded as a clear test of realism. Next, we proceed to discuss the specifics of our example.

\textit{LGI using the LHO Schr\"{o}dinger coherent state}: Let us consider the following initial Gaussian wave function 

\begin{equation}
\label{LG3}
\psi(x,t=0)=\sqrt{\frac{1}{\sqrt{2\pi}\sigma_{0}}} \exp{\left(-\frac{x^{2}}{4 \sigma^{2}_{0}}+\frac{ip_{0} x}{\hslash}\right)}
\end{equation}

\noindent with the initial momentum expectation value $p_{0}$, and the width $\sigma_{0} = \sqrt{\frac{\hbar}{2m\omega}}$ where $\omega$ is the angular frequency of oscillation. It is well known that under the LHO potential, the above $\psi(x,0)$ evolves into $\psi(x,t)$ (whose detailed expression is given in the Supplementary Material I \cite{sup}), whence the probability density is given by 

\begin{equation}
\label{LG4}
\vert\psi(x,t)\vert^{2}=\sqrt{\frac{m\omega}{\hslash\pi}}\exp{\left(-m\omega\frac{(x-\frac{p_{0}}{m\omega}\sin \omega t)^{2}}{\hslash}\right)}
\end{equation}

\noindent which oscillates without spreading or changing shape, while its peak follows classical motion, and $\Delta x \Delta p = \hbar/2$ at all instants. Such a wave packet is known as the Schr\"{o}dinger coherent state \cite{schrodinger} - a much-discussed remarkable example of a quasi-classical state in quantum mechanics. In order to apply LGI in this context, we consider coarse-grained measurement of a type that determines at any instant whether the oscillating particle is in the region between $x\rightarrow -\infty$ and $x=0$ (yielding the measurement outcome +1) \textit{or} is in the region between $x=0$ and $x\rightarrow +\infty$  (yielding the measurement outcome -1). Such a measurement can be represented by the localization operator $\hat{O}=\int_{-\infty}^{0}\vert x\rangle\langle x\vert dx -\int_{0}^{\infty}\vert x\rangle\langle x\vert dx$ which has two eigenstates $\int_{-\infty}^{0}\langle x|\psi\rangle\vert x \rangle dx$ and $\int_{0}^{\infty}\langle x|\psi\rangle\vert x \rangle dx$ corresponding to the eigenvalues $+1, -1$ respectively. We will later comment on the feasibility of measuring an operator close to $\hat{O}$, making the point that an ideal sharp boundary at $x=0$ for distinguishing the $+1$ and $-1$ outcomes is not really required.  Now, note that the probability of obtaining the outcome +1(-1) for such a measurement at the instant, say, $t_1$, is given by 

\begin{equation}
\label{LGI5}
P_{+}(t_{1})=\int_{-\infty}^{0}\vert\psi(x,t)\vert^{2} dx=\frac{1}{2}\left(1-\mathrm{\mathrm{erf}}(\frac{\langle x(t)\rangle}{\sqrt{2}\vert\sigma_{t}\vert})\right)
\end{equation}

\begin{equation}
\label{LGI6}
P_{-}(t_{1})=\int_{0}^{\infty}\vert\psi(x,t)\vert^{2} dx=\frac{1}{2}\left(1+\mathrm{erf}(\frac{\langle x(t)\rangle}{\sqrt{2}\vert\sigma_{t}\vert})\right)
\end{equation}

\noindent where the Error Function $\mathrm{erf}(t)=\frac{2}{\sqrt{\pi}}\int_{0}^{t} \exp(-z^{2})dz$ and $\sigma_{t}=(i\hbar \sin\omega t+ 2m\omega\sigma_{0}^{2} \cos \omega t)/2m\omega \sigma_{0}$.

Next, given the result of the above measurement at the instant $t_{1}$ to be +1(-1), obtained using the NRM procedure (its suggested empirical implementation in this case is discussed later), the subsequent time evolution of the post-measurement state is subjected to a measurement at an instant, say, $t_{2}$. For this latter measurement, the conditional probability of obtaining the outcome +1, contingent upon the outcome +1(-1) obtained for the measurement at the earlier instant $t_{1}$, is given by

\begin{equation}
\label{LGI9}
P_{\pm/+}(t_{1}, t_{2})=\int_{-\infty}^{0}\vert\psi_{\pm}^{PM}(x,t_{2})\vert^{2} dx
\end{equation}
                                                                       
\noindent while such a conditional probability for the outcome -1 at the instant $t_{2}$ is of the form

\begin{equation}
\label{LGI10}
P_{\pm/-}(t_{1}, t_{2})=\int_{0}^{\infty}\vert\psi_{\pm}^{PM}(x,t_{2})\vert^{2} dx
\end{equation}

\noindent where $\psi_{\pm}^{PM}(x,t_{2})$ is the time-evolved form of the post-measurement state that has evolved up to the instant $t_{2}$, and whose expression is given in the Supplementary Material II \cite{sup1}. 

\textit{Results}: Using Eqs. (\ref{LGI5}) - (\ref{LGI10}), for suitable choices of the relevant parameters, one can compute the QM values of the joint probabilities $P_{++}(t_{1}, t_{2})$, $P_{+-}(t_{1}, t_{2})$, $P_{--}(t_{1}, t_{2})$, $P_{-+}(t_{1}, t_{2})$ and evaluate the temporal correlation function $C_{12}$. Similarly, the other temporal correlation functions $C_{23}$,$C_{34}$,$C_{14}$ occurring in LGI of the form (1) can be calculated. In our setup, the key parameters are m, $p_{0}$ and $\omega$. Suitably choosing the values of m, $p_{0}, \omega$ while taking the temporal intervals to be the same, i.e., $t_{2}- t_{1}=t_{3}- t_{2}=t_{4}- t_{3}=\Delta t$, and by numerically integrating the relevant integrals occurring in Eqs.(\ref{LGI5}) - (\ref{LGI10}), the key results of the quantitative studies are presented in the Tables I - III. Here it needs to be mentioned that for given values of m, $p_{0}$ and $\omega$, by varying the choices of the time interval $\Delta t$ and the first instant of measurement $t_{1}$, it is found that the maximum value of C on the LHS of the inequality (1) is attained when $\Delta t$ is chosen within the neighbourhood of T/4 or 3T/4, and $t_{1}$ is slightly larger than 0 or is within the neighbourhood of T/2, where T is the time period of oscillation. Note that for computing all the results given in the Tables I - III, we have chosen the \textit{same} values of $\Delta t = 2.4 \times 10^{-6}$s and $t_{1}=1.5 \times 10^{-6}$s where $\Delta t$ is chosen close to 3T/4 and $t_{1}$ is close to T/2 with $T=3.14 \times 10^{-6}$s which corresponds to $\omega = 2 \times 10^{6}$Hz (this value of $\omega$ is close to the typical value $100$KHz-$1$MHz for optically levitated oscillating masses). We now proceed to summarise below the results given in the Tables I-III:   

\begin{table}
\label{tab1}
\caption{\footnotesize {Taking the angular frequency of oscillation $\omega = 2 \times 10^{6}$Hz, for various values of mass (m), different choices of the initial peak momentum $p_{0}$ (initial peak velocity $v_{0}$) of the coherent state wave packet are indicated for which the respective QM values of the LHS (C) of the LGI inequality(1) are computed. The corresponding values of the constant width ($\sigma_{0}$) of the coherent state wave packet and the classical amplitude ($A_{Cl}$) of oscillation are given.}}
\begin{tabular}{|c|c|c|c|c|c|}
\hline 
m(amu) &$ \sigma_{0} (m) $ & $p_{0}(kgm/s) $ & $v_{0}(m/s)$ & $ A_{Cl}(m) $ &C \\  
\hline
$10$ &  $3.9\times10^{-8}$ & $ 3.3\times 10^{-24}$ &  $2 \times 10^{2}$ & $ 10^{-4}$ & 2.62 \\ 
\hline
$10^{3}$ &  $3.9\times10^{-9}$ & $ 3.3\times10^{-23}$   &$2 \times 10$&  $ 10^{-5}$ & 2.58 \\ 
\hline 
$10^{6}$ &  $1.2\times10^{-10}$ & $ 3.3\times10^{-21}$  &2.0 & $ 10^{-6}$& 2.5 \\ 
\hline
$10^{10}$ &  $1.2\times 10^{-12}$ &  $ 3.3\times10^{-21}$ & $2\times 10^{-4} $ & $ 10^{-10}$& 2.7 \\ 
\hline
$10^{20}$ &  $1.2\times 10^{-17}$ &  $ 3.3\times 10^{-15}$ & $2\times 10^{-8} $ & $ 10^{-14}$ & 2.65 \\ 
\hline
\end{tabular} 
\end{table}

\begin{table}
\label{tab2}
\caption{\footnotesize{Taking fixed values of $\omega = 2 \times 10^{6}$ Hz and $p_{0}=3.3 \times 10^{-24}$kgm/s, for increasing values of m, gradual decrease of the QM violation of LGI is shown through decreasing values of C, while the corresponding values of $\sigma_{0}$, $v_{0}$, $A_{Cl}$ are indicated.}}
\begin{tabular}{|c|c|c|c|c|}
\hline  m(amu) & $\sigma_{0}(m)$& $v_{0}(m/s)$ & $A_{Cl}(m)$ & C\\ 
\hline  $10^{2}$ & $1.2\times 10^{-8}$& $2 \times 10^{2}$ & $10^{-5}$ & 2.8\\ 
\hline  $10^{3}$ & $3.8\times 10^{-9}$& 2.0 & $ 10^{-6}$ & 2.74\\ 
\hline  $10^{4}$ & $1.2\times 10^{-9}$&$2 \times 10^{-1}$ & $ 10^{-7}$& 2.65\\ 
\hline  $10^{6}$ & $1.2\times 10^{-10}$&$10^{-3}$ & $10^{-9}$& 1.56\\   
\hline
\end{tabular}
\end{table}

\begin{table}
\label{tab3}
\caption{\footnotesize{Taking fixed values of $m=10^{3}$amu and $\omega = 2 \times 10^6$Hz,, for increasing values of $p_{0}$ that correspond to increasing values of $A_{Cl}$, the respective computed QM values of the LHS (C) of the LGI inequality (1) are shown which indicate a gradual decrease in the QM violation of LGI as the value of $A_{Cl}$ increases, and eventually LGI is satisfied.}}
\begin{tabular}{|c|c|c|c|c|c|}
\hline  m(amu) &$p_{0}(kgm/s)$&$v_{0}(m/s)$& $\sigma_{0}(m)$ & $A_{Cl}(m)$ & C\\ 
\hline  $10^{3}$ & $3.32\times 10^{-25} $&$2 \times 10^{-1}$ & $3.9\times 10^{-9}$  & $10^{-7}$ & 2.54\\ 
\hline  & $3.32\times 10^{-24}$& 2 &$3.9\times 10^{-9}$ &$10^{-6}$ & 2.73\\ 
\hline   &$3.32\times 10^{-23}$& $2 \times 10$ &$3.9\times 10^{-9}$& $ 10^{-5}$& 2.6\\ 
\hline  & $3.32\times 10^{-22}$& $2 \times 10^{2}$ &$3.9\times 10^{-9}$ &$10^{-4}$& 1.99\\  
\hline
\end{tabular}
\end{table}

(a) It is found that while for the peak momentum $p_{0}=0$, LGI is always satisfied, by appropriately choosing $p_{0}$, it is  possible to obtain a significant amount of QM violation of LGI for any m corresponding to a given $\omega$, and this violation can be maximised over $\Delta t$ and $t_{1}$ (as per the choices of $\Delta t$ and $t_{1}$ mentioned above). This is illustrated by the results presented in Table I where the maximum obtained values of C are given for different sets of values of the relevant parameters, corresponding to a given $\omega = 2 \times 10^{6}$ Hz while the mass is varied from 10 amu to $10^{20}$ amu. Note that appreciable QM violations of LGI $(C>2)$ are found by suitable choices of $p_{0}$ given in Table I for, say, masses 10 amu $-10^{10}$ amu, such that the respective values of the classical amplitude of oscillation $A_{Cl}$ $= p_0/m\omega$ range from $10^{-4}$m to $10^{-10}$m. If the mass is further increased to, say, $10^{20}$ amu, it is found that in order to obtain significant QM violation of LGI, $p_{0}$ needs to be chosen such that the corresponding $A_{Cl}$ becomes much smaller. Also, as m increases, the required value of $v_{0}$ (initial peak velocity of the wave packet) for showing the QM violation of LGI becomes increasingly smaller. Thus, although theoretically one can obtain the QM violation of LGI for any given m and $\omega$ by suitably choosing $p_{0}$, actual testability of this violation becomes gradually impracticable for sufficiently large mass as the requirement to controllably impart exactly the appropriate momentum becomes more stringent.

(b) The results given in Table II show that if by keeping the parameters $p_{0}, \omega$ fixed, one increases the mass m, the QM violation of LGI gradually diminishes, and eventually for sufficiently large mass, LGI is satisfied; i.e., $C<2$. 

(c) For given values of m and $\omega$, if $p_{0}$ is increased, the corresponding $A_{Cl}$ is also increased, the QM value of C is found to be gradually decreasing, and eventually $C<2$ for appropriately large $p_{0}$. This is illustrated by Table III.

The results discussed above, therefore, serve to highlight the efficacy of LGI in not only revealing a nonclassicality of the oscillator coherent state, but also in exploring the extent to which such nonclassical feature persists for masses larger than the typical microsopic masses. A preliminary sketch of a realizable setup that can verify the above predicted results is now indicated.

\textit{A proposed experimental scheme}: We consider two illustrative cases: (i) a nano-object of $10^6$ amu  trapped by laser fields \cite{Ashkin} that generate a harmonic well of $\omega \sim$MHz \cite{Ulbricht} (from Table I, it is seen that C=2.5 with $A_{Cl}\sim \mu$m) and (ii) an ionized nano-object of $10^{9}$ amu trapped in an ion-trap of $\omega \sim$100 Hz \cite{Millen} (for this case we have estimated that C=2.7 for $A_{Cl}\sim \mu$m). Damping and decoherence in both the cases are negligible in the experimental time-scale of $1/\omega$ so that the time evolution is well approximated by the unitary dynamics as used in our treatment \cite{Scala,ion-trap-heating}. A primary criterion is to be able to differentiate reasonably sharply the presence of the mass on the left or on the right halves of the harmonic well in implementing the measurement of the operator $\hat{O}$. The positions of the optically levitated masses have already been observed with extremely high spatial resolution by means of photo-diodes using the interferometric (phase sensitive) detection of light scattered from the objects \cite{Raizen,Novotny,Ulbricht-private}. This technique enables the detection of positions of the aforementioned masses in their corresponding traps with sub-Angstrom resolutions much sharper than the spread $\sigma_{0}$. Moreover, the detecting time-window to achieve this resolution is much smaller than $1/\omega$ so that the measurements can essentially be regarded as instantaneous. By criss-crossing the $x>0$ part of the well with several such scattering light fields whose intensity fall to zero sharply at $x=0$, in case the nano-object fails to scatter light (i.e., is essentially invisible), then its state is projected to the eigenstate of $\hat{O}$ with eigenvalue $-1$ (this corresponds to the NRM procedure as no light has interacted with the nano-object for yielding this outcome). Note that the QM violation of LGI is retained for a significant unsharpness of dichotmic measurements \cite{Mal} so that even if one limits the recourse to measurement precision coarser than the state of art, then resolutions of up to $\sim 0.2\sigma_0$ (in conformity with the robustness of the QM violation of LGI with respect to the extent of unsharpness of measurement) should suffice for observing the LGI violation in our setup. As regards preparing the initial state, the nano-object can at first be laser cooled to its ground state (i.e.,the Gaussian wavefunction $\psi(x,t=0)$ with $p_0=0$) even at room temperatures by using the schemes \cite{barker,romero-isart,zoller} which are tantalizingly close to being achieved \cite{Kiesel,Asenbaum,Millen}. Subsequently, the centre of the trap can be suddenly displaced (as demonstrated in ion traps \cite{Wineland}) so that the wavepacket is centred at the $A_{Cl}\sim \mu$m. Then, after a quarter oscillation, the wavepacket can gain the appropriate momentum $p_0$ for the LGI violation.

\textit{Concluding remarks}: We have shown how a system having a well defined classical limit, namely the quantum harmonic oscillator, can be made to violate LGI through suitable spatial measurements even when the initial state is the most classical of all states -- namely the Schr\"{o}dinger coherent state. The distinctive positive features of two recent proposals: the usage of a harmonic oscillator \cite{Rabl} and of a direct position measurement of an object \cite{rob} are both invoked in our scheme, while ours is qualitatively very different from either (for example, neither ancillary qubits, nor a tunneling between distinct potential wells are involved). As already explained, the dichotomic spatial measurement used in our example is feasible to a very good approximation. This measurement, which projects the system to a highly non-classical state, gives rise to the LGI violation and provides arguably the most convincing  NRM possible as outcomes are retained essentially when the measurement probe has no interaction whatsoever with the system under consideration. Strikingly, the better the fine control one can acquire on trap-displacements or momenta, the larger the mass for which LGI violation can be observed, thus offering an avenue for extending the test of the limits of quantum behaviour to the macroscopic domain. In practice, larger $m$ will require lower $\omega$ in order to have $\sigma_0$ larger than the feasible precision of position measurements, and this will in turn extend the time-scale of the experiment. When this would exceed the typical decoherence time ($1-10$ms for levitated objects), our calculations have to be modified to include the decoherence effects. 

To summarise, the above discussed research programme involving a system like oscillator which has a familiar classical analogue in our everyday world should be a worthwhile contribution to the line of studies that seeks to probe the macro-limits of the quantum world in conjunction with the testing of the notion of realism. Finally, since the LGI violation for an \textit{isolated} oscillator is in itself yet unexplored, this should be worth testing even in the micro-domain with trapped ions, using electromagnetic fields in cavity and circuit-QED.

\textit{Acknowledgements}: We thank Peter Barker, Anis Rahman, Hendrik Ulbricht and Nikolai Kiesel for answering our queries on the position detection of levitated nano-particles. Thanks are also due to G. S. Agarwal and N. D. Hari Dass for helpful comments on an earlier version of this work. SB would like to acknowledge the Engineering and Physical Sciences Research Council Grant No. EP/J014664/1. DH and SM acknowledge the support provided by the Dept. of Science and Technology, Govt. of India.

\pagebreak
\widetext
\begin{center}
\textbf{\large Supplemental Materials: Uncovering a Nonclassicality of the Schr\"{o}dinger Coherent State up to the Macro-Domain}
\end{center}

\setcounter{equation}{0}
\setcounter{figure}{0}
\setcounter{table}{0}
\makeatletter
\renewcommand{\theequation}{S\arabic{equation}}
\renewcommand{\thefigure}{S\arabic{figure}}
\renewcommand{\bibnumfmt}[1]{[S#1]}
\renewcommand{\citenumfont}[1]{S#1}

We provide here  expressions for the time-evolved form of the coherent state wave packet used in our paper, as well as for the post-measurement state arising from the coarse-grained measurement of the type described in the paper.

\begin{center}
\textbf{Supplement I}

\vskip 0.2cm 

{\it\textbf{Expression for the time-evolved coherent state wave packet}}\\
\end{center}
The coherent state wave packet at t=0 given by Eq. (3) in the text, with $\sigma_{0} = \sqrt{\frac{\hbar}{2m\omega}}$, is evolved in the linear harmonic potential by the following propagator

\begin{eqnarray}
K\left(x^{\prime}, t^{\prime} = 0;x,t\right)=\sqrt{\frac{m\omega}{2\pi i \hslash\sin\omega t}} \exp\left[\frac{i m\omega}{2\hslash\sin\omega t} \{(x^{\prime 2}+x^{2})\cos\omega t-2x x^{\prime}\}\right]
\end{eqnarray}

Then the time-evolved wave packet at the instant t is given by

\begin{eqnarray}
\psi(x,t)=\int_{-\infty}^{\infty} K(x^{\prime},t^{\prime}=0;x,t)\psi(x^{\prime},0) dx^{\prime}\nonumber\\
=\sqrt{\frac{1}{\sqrt{2\pi}\sigma_{t}}}\exp{(-\sqrt{m\omega}\frac{A(t)+B x+C(t)x^2}{(2\hslash)^{3/2}\sigma_{t}}}).
\end{eqnarray} \\

\noindent where 
\begin{eqnarray}
A(t)=\frac{i\hslash p_{0}^{2}}{(m\omega)^{2}}\sin\omega t\\
B=-\frac{2 i p_{0}\hslash}{m\omega}\\
C(t)=\hslash\cos\omega t+ i \hslash\sin\omega t\\
\sigma_{t}=\frac{i\sin\omega t+ \cos\omega t}{\sqrt{2 m\omega/\hslash}.}
\end{eqnarray}\\

\begin{center}
\textbf{Supplement II}

\vskip 0.3cm

{\it\textbf{Expressions for the post-measurement state and its time-evolved form}}\\
\end{center}
Depending on the outcome $+1(-1)$ of the measurement corresponding to the operator $\widehat{O}=\int_{-\infty}^{0}\vert x\rangle\langle x\vert dx -\int_{0}^{\infty}\vert x\rangle\langle x\vert dx$ at the instant $t_{1}$, the post-measurement state (not normalized) is given by

\begin{eqnarray}
\vert\psi_{+}^{PM}(t_{1})\rangle= \int_{-\infty}^{0} \psi(x^{\prime},t_{1})\vert x^{\prime}\rangle dx^{\prime}\\
\vert\psi_{-}^{PM}(t_{1})\rangle=\int_{0}^{+\infty}\psi(x^{\prime},t_{1})\vert x^{\prime}\rangle dx^{\prime}.  
\end{eqnarray}\\

Subsequently, $\psi_{\pm}^{PM}(t_{1})$ evolves up to the instant $t_{2}$ by the propagator $K(x^{\prime}, t^{\prime}=t_{1}; x, t_{2})$ which is of the same form as that given by Eq. (1). The time-evolved normalized form of the post-measurement state at the instant $t_{2}$ is then given by

\begin{eqnarray}{\nonumber}
{\psi}_{\pm}^{PM}(x,t_{2})=(1/N_{\pm})\int_{-\infty}^{\infty} K(x^{\prime},t_{1};x,t_{2})\psi_{\pm}^{PM}(x^{\prime},t_{1}) dx^{\prime}
\end{eqnarray}
\begin{eqnarray}
= (1/N_{\pm}) \frac{1}{2\sqrt{\sqrt{2\pi}\sigma_{t_{2}}}}(1+Erf[\frac{\chi}{\sqrt{\xi}}])\exp\left[-\sqrt{m\omega}\frac{A(t_{2})+B x+C(t_{2})x^{2}}{(2\hslash)^{3/2}\sigma_{t_{2}}}\right]
\end{eqnarray}\\

\noindent where $A(t_{2}), B, C(t_{2})$, and $\sigma_{t_{2}}$ are respectively the same as that given by Eqs. (3), (4), (5) and (6), except that t is replaced by $t_{2}$, while

\begin{eqnarray}
\chi=-\frac{\sqrt{m\omega}B}{(2\hslash)^{3/2}\sigma_{t_{2}}}-\frac{i m\omega x}{2\hslash\sin\omega(t_{2}-t_{1})}\\
\xi=\frac{\sqrt{m\omega}C(t_{2})}{(2\hslash)^{3/2}\sigma_{t_{2}}}-\frac{i m\omega\cos\omega (t_{2}-t_{1})}{2\hslash\sin\omega(t_{2}-t_{1})}
\end{eqnarray}\\

\noindent and $N_{\pm}$ is the normalisation constant at the instant $t_{2}$, given by

\begin{eqnarray}
N_{\pm}=\int_{-\infty}^{\infty}\vert\psi_{\pm}^{PM}(x,t_{2})\vert^{2} dx
\end{eqnarray}

\end{document}